MANUSCRIPT

# Pan-genome Analysis of the Genus *Serratia*

**Zarrin Basharat*, Azra Yasmin**

Microbiology & Biotechnology Research Lab, Department of Environmental Sciences, Fatima Jinnah Women University, Rawalpindi 46000, Pakistan.

***Corresponding author**
**Email:** zarrin.iiui@gmail.com

**ABSTRACT**
**Pan-genome analysis is a standard procedure to decipher genome heterogeneity and diversification of bacterial species. Specie evolution is traced by defining and comparing the core (conserved), accessory (dispensable) and unique (strain-specific) gene pool with other strains of interest. Here, we present pan-genome analysis of the genus *Serratia*, comprising of a dataset of 100 genomes. The isolates have clinical to environmental origin and consist of ten different species from the genus, along with two subspecies of the representative strain *Serratia marcescens*. Out of 19430 non-redundant coding DNA sequences (CDS) from the dataset, 972 (5%) belonged to the core genome. Majority of these genes were linked to metabolic function, followed by cellular processes/signalling, information storage/processing while rest of them were poorly characterized. 10,135 CDSs (52.16%) were associated with dispensible genome while 8,321 CDSs (42.82%) were singletons or strain specific. The Pan-genome orthologs indicated a positive correlation to the number of genomes whereas negative correlation was obtained for core genome. Genomes were aligned to obtain information about synteny, insertion/inversion, deletion and duplications. This study provides insights into variation of *Serratia* species and paves way for pan-genome analysis of other bacterial species at genus level.**

## 1. INTRODUCTION

Genus *Serratia* consists of rod-shaped, Gram-negative, endospore forming, facultative anaerobic bacteria belonging to the family Enterobacteriaceae (Li et al., 2015). They exist in diverse habitats including soil (Kamensky et al., 2003; Giri et al., 2004; Lavania and Nautiyal, 2013), water (Gavini et al., 1979; Henriques et al., 2013; Kämpfer and Glaeser, 2016), plants (Lim et al., 2015; Afzal et al., 2016; Zaheer et al., 2016), nematodes (Abebe-Akele et al., 2015; Vicente et al., 2016), bats (García-Fraile et al., 2015), humans (Roy et al., 2014; Bonnin et al., 2015) etc, and consist of non-pathogenic as well as some pathogenic strains.

Some *Serratia* species stimulate plant growth through production of specific compounds as siderophores, lipo-chitin, oligosaccharides etc and are considered plant growth promoting rhizobacteria (Koo and Cho, 2009; Purushotham et al., 2012). Their antagonistic behaviour against phytopathogens has been documented due to a diverse array of antimicrobial compound production (De Vleesschauwer and Hofte, 2003; Adam et al., 2016; Roberts et al., 2016). Prodiginine (with five sub-types i.e. prodigiosin, undecylprodigiosin, cycloprodigiosin, cyclononylprodigiosin, butyl-meta-cyclo-heptylprodiginine) is the most well characterized antimicrobial compound in *Serratia* (Williamson et al., 2006). *Serratia marcescens*, a typical/representative strain is an opportunistic pathogen and causes nosocomial infections (Gerc et al., 2015). However, some reports of infections by *Serratia plymuthica, Serratia liquefaciens, Serratia rubidaea*, and *Serratia odoriferae* also exist (Weise et al., 2014) Members of *Serratia marcescens* are distinguished from other Enterobacteriaceae family members due to production of enzymes DNase, lipase, and gelatinase which are not produced by the other species (De Vleesschauwer and Hofte, 2003; Li et al., 2015). Prodigiosin, a red pigment with antibacterial, antifungal, antiprotozoal, immunosuppressive and anticancer properties (Williamson et al., 2007) is normally produced by environmental isolates of *Serratia marcescens*, but not the clinical isolates (Mahlen, 2011).

High throughput methods for genome sequencing with reduced time and cost have resulted in massive DNA-sequencing data in recent years (Mardis, 2008; Goodwin et al., 2016). To encompass the detailed pan genomic repertoire of genus *Serratia,* a hundred genomes were taken, encoding for all possible lifestyles carried out by the species. The phylogenetic resolution of the genus of interest ranged from species to subspecies/serovar. This effort was centred on heterogeneity analysis and evolutionary study of this genus.

## 2. MATERIAL AND METHODS

Genomes used in this study were obtained from the public repository 'NCBI database' (Table 1). Statistical parameters were calculated for the genomes such as genome size in Mbp (million base pairs), DNA GC content in percentage, Number of CDSs, Mean length of CDSs etc. Pairwise analysis was done via ortholog matrix (parameters: Reference Genome = *Serratia marcescens* subsp. *marcescens* Db11, Data type = RBH protein) (Manzano-Marin et al., 2012) as previously implemented for seven *Serratia* strains. while whole genome alignment was carried out using NUCMER (Delcher et al., 2002; Kurtz et al., 2004; Hasan et al., 2015), to reveal differences caused by duplication, inversion, deletion and insertion.

The strain phylogeny was inferred through orthologous average nucleotide identity (OrthoANI) algorithm (Lee et al., 2016). Gene frequency and the gene presence/absence heatmap was constructed for all 100 genomes through UPGMA clustering (Comandatore et al., 2013).

A pan-genome is a pool or amalgamation of protein coding genes i.e. CDSs in a specified set of genomes (Vernikos et al., 2015). Set of annotated genomes are taken as a seed substance for the construction of a pan-genome. Genomes are subjected to a pair wise homology search through BLAST or uBLAST. "Reciprocally best hit (RBH)" orthologs are calculated for all possible genome pairs. In case of a partial gene, the reciprocity cannot be defined owing to small length. The gene, therefore, is considered as a singleton/different gene, even though it may have 100% sequence similarity with the orthologous group gene. uCLUST was implemented to circumvent this problem, and genes were regrouped in case of ~95% identity according to Cao et al. (2013).

Pan-genome information was then converted into binary matrix based on the presence (coded as 1) and absence (coded as 0) of genes (Laing et al., 2010; Georgiades et al., 2011). Similarity values were then calculated between two genomes/POGs based on Jaccard coefficient. The obtained similarity matrix was then used for clustering potential CDSs via UPGMA algorithm into non-redundant gene sets, to generate POGs (Contreras-Moreira and Vinuesa, 2013). Frequency of POGs (Lefébure and Stanhope, 2007), functional categorization through COG (Altenhoff and Dessimoz, 2009) and SEED scheme (Disz et al., 2010) was obtained. The pan/core genome curves were then calculated from POGs (Zakham et al., 2012).

## 3. RESULTS AND DISCUSSION

Similar and dissimilar properties of the genomes were matched through statistical properties of genome size in Mbp (million base pairs), DNA GC content in percentage, Number of CDSs, Mean length of CDSs (Table 2) and simple regression analysis (Table 3). Remarkable differences and similarities were observed.

Alignment of two whole genome sequences can be very useful in comparative genomics and it is particularly helpful when alignments are visualized. Gene synteny and genomic structural difference caused by duplication, inversion, deletion and insertion are inferred by this way. For a dot plot, the reference sequence is laid across the x-axis, while the query sequence is on the y-axis. Wherever the two sequences agree, a coloured line or dot is plotted. The forward matches are displayed in red, while the reverse matches are displayed in green. If the two sequences are perfectly identical, a single red line would go from the bottom left to the top right. However, two sequences seldom exhibited this behaviour, and in the studied plots, multiple gaps and inversions were identified for majority genomes. *Serratia plymuthica* AS12, *Serratia liquefaciens* ATCC 27592, *Serratia nematodiphila* WW4 and *Serratia proteamaculans* S4 showed a significance degree of similarity to the reference strain *Serratia marcescens* subsp. *marcescens* Db11 whereas least similarity was observed for *Serratia grinesii* A2, Serratia subsp *sakuensis* and *Serratia symbiotica* SAf with respect to the reference strain (Fig. 1).

ANI based typing differed for several strains from the traditional typed strain names. An example is of *Serratia marcescens* (NCBI project accession number: GCF_000336425.1) demarcated as *Serratia nematodiphila*. ANI and pan-genome studies often challenge the traditionally typed bacterial strain names (Kim et al., 2014; Yi et al., 2014). In this case too, several bacteria were reannotated as new species. A phylogenetic tree was constructed based on ANI and genomes with similar content were clustered together (Fig. 2).

The genome content and genes within the studied bacterial genomes might be altered due to environmental stress, horizontal gene transfer etc. Despite continual variation in whole genome content of bacteria, orthologous genes are shared between multiple genomes. In case of evolutionarily related bacteria, more genes are shared while in case of distantly related species, number of common genes is less. Several strains showed varied pan genomic profile due to presence of significant number of genes that were absent in the other strains and were clustered near to each other depending on the type and number of these genes (Fig. 3). A significant portion of the whole genome (for all studied genomes) belonged to the house keeping/core genes required for essential functions and were present in all of the species while some genes (called singletons) are present only in a single genome. These singletons usually belong to the genomic islands/mobilome with horizontal gene transfer, conjugation, transformation or transduction origin. A "gene frequency plot" helps distinguish the varying trend. A U-shaped plot is usually obtained when most genes are present either in all or just single genome. A U-shaped plot was obtained for the genes of *Serratia* species.

After formulating pan-genome, all CDSs were grouped into pan-genome orthologous groups (POGs). POGs enclosed a minimum of one CDS (called singleton POGs). Extremely conserved POGs were present in all 100 genomes, and these constituted the core genome. For 100 genomes of *Serratia*, 19,430 POGs (~4%) were obtained from 485,201 CDSs for pan-genome and 972 POGs (~5%) were obtained from 485,201 CDSs for the core region. Previously, a comparative study by Li et al. (2015) reported that ~41% of the predicted CDSs for pan-genome were shared among ten *Serratia* genomes (Li et al., 2015). This number was reduced to 4% for 100 genomes. This illustrates the need for increased number of genomes for determining pan and core genome instead of just a few genomes. It helps elucidate the conserved and varying regions in a better way. Some notable examples of bacteria which appear to have an open pan-genome include *E. coli* (Collins and Higgs, 2012), *Streptococcus agalactiae* (Tettelin et al., 2005), *Prochlorococcus* (Kettler et al., 2007), *Listeria* (den Bakker et al. 2010) etc whereas closed pan-genome has been deciphered for *Clostridium difficile* (Scaria et al., 2010), with a sample size of 26 genomes rendered sufficient for discernment of the entire pan-genome. However, this result may depend on the way the extrapolation was done and may not indicate a qualitative difference between *Clostridium difficile* and other bacteria. Expanding the idea of open and close pan-genome, it should be noted that mostly environmental bacteria have an open pan-genome while other bacteria exhibit closed pan-genome (Vernikos et al., 2015), but at genus scale, bacteria from diverse type of habitats and varying characteristics are included and it is difficult to comment upon the closedness or openness of its pan-genome. Vernikos et al. (2015) have suggested a combinatorial method for pan-genome prediction, where average values are taken for all data types in the set. This is attempted by first calculating the new, core and shared genes at the nth genome through the following function, with C being the total number of comparisons and N total number of genomes.

$$C = N!/(n-1)!.(N-n)!$$

Scalability issue has been addressed by developing a method for sub-sampling the number of comparisons to be performed between N genomes through random selection and comparison. Each strain is however, ensured to undergo equal number of comparisons. Some additions in the metric are required which might account for the environmental/pathogenic isolates, habitat, genome size, total number of genes, threshold for calculation of ortholog relatedness and related parameters for calculation of pan-genome openness or closedness. We postulate that a pan-genome should always be open in bacteria because the genomes are continuously evolving, either due to horizontal gene transfer, environmental impact or new gene arisal or deletion simply according to the neutral Wright–Fisher model (Wright 1969; Baumdicker et al., 2010)

Frequency of POGs, functional categorization through COG and SEED scheme was studied which illustrated a visible difference between pan and core genome for specific set of proteins (related to particular cell functions) for all studied genomes (Fig. 4). Functional categories according to SEED and COG with a cut-off of 50% and 100% were studied and variation of core genes to the pan genome was highly contrasting. The contrasting categories according to SEED scheme included DNA/RNA/protein coding machinery and co-factor/vitamin/prosthetic group genes as well pigment synthesis genes more localized within core genome region. A small portion of the core while a large portion of the pan-genome demonstrated no match to the database contents. According to COG functional categorization, amino acid transport/metabolism, translational machinery, ribosomal structure proteins, chaperones and post translational modification apparatus contributed significantly to the core portion as compared to the miniscule portion of these genes in the pan-genome region. Only a tiny portion of core genome with a cut-off less than 100% and pan-genome regions could not be related to the database contents. POGs (shown as boxplots) signify the distribution of values of genome data set obtained through random selection of the genomes on X-axis, 100 times each. Boxplot and trend line illustrations the shift in number of pan-genome and core genome (Fig. 5) gene groups to the number of genomes. The pan-genome curve shows a positive correlation between the number of POGs and the number of genomes. Pan-genome analysis of 100 genomes with the cut-off setting at 100% resulted in core genome that included only the genomes with all of the certain genes. In contrast, the core genome curve showed a negative correlation where the number of POGs decreased as the number of genomes increased. In the case of core genome, it was calculated by using a cut-off setting of 80 and 85 and 100%. The cut-off setting specified the calculation of core genome by using genomes with certain genes that were above 80, 85 and 100% of the genomes in the entire data set. Overall, we saw that the number of POGs that belonged to the core-genome increased as the cut-off percentage got smaller.
Core-genome size of this bacteria is small (972 genes for ~5Mbp/100 genomes) as compared to *Campylobacter jejuni* (1042 genes for ~1.6 Mbp/130 genomes) (Meric *et al.*, 2014) but sufficiently large as compared to the *Bacillus* species (143 genes for ~4.5 Mbp/172 genomes) (Collins and Higgs, 2012). The *Bacillus* specie might have reduced core as number of genomes were large and they are different from their own subspecie i.e. *Geobacillus* (527 genes for ~4.5 Mbp/29 genomes). This might be due to small sample size in case of *Geobacillus* (Bezuidt et al., 2016). Overall, this observation is a bit interesting and warrants further investigation. Codon usage bias analysis and fitting of pan-genome data using various models like random permutation model, power law regression model, exponential regression model etc would also be of aid in better understanding evolution of bacteria. The authors are currently involved in studying phage and mobilome impact on pan-genomics of *Serratia*. We expect that similar efforts at genus scale for other bacteria would also be undertaken by our group in the future.

## 4. CONCLUSION

We have presented the pan-genome analysis of *Serratia* species and this has allowed us to gain an improved understanding of this versatile genus. Individual genomes of the *Serratia* are made up of core and pan CDS. Mining the accessory genomes and alignment of species has shed light on the genome conservation. Compared to other bacterial species, *Serratia* does not depict exceptional genetic conservation or diversity. Standardization of pan-genome parameters would lead to understanding the details of this and contribute to knowledge of different phenotypes compared to molecular genotypes. Incorporation of expression data, metabolic pathways and mobile elements etc is also desired. With the development of new approaches and better tools on horizon, future of bacterial pan-genomics seems bright.

**NOTE: Supplementary data consisting of core and pan genome orthologs is available upon request from the authors.**

## ACKNOWLEDGEMENTS
The authors are thankful to Alexandra Elbakyan for providing literature support. Furthermore, we are grateful to all those scientists, research labs and universities who made computational biology tools, data and web servers freely available for public use.

## FUNDING
This study did not receive any particular funding.

## CONFLICT OF INTEREST
The authors declare that they have nothing to disclose which might pose a potential conflict of interest.

Table 1. Characteristics of the studied 100 *Serratia* genome species.

| Serial no. | NCBI project Accession no. | Strain name | Name | Status | No. of contigs | Genome size (bp) | DNA G+C content (%) | No. of CDSs | No. of rRNA genes | No. of tRNA genes |
|---|---|---|---|---|---|---|---|---|---|---|
| 1 | GCF_000018085.1 | 568 | *Serratia* proteamaculans 568 | COMPLETE | 2 | 5495657 | 55.02383 | 5066 | 22 | 85 |
| 2 | GCF_000214195.1 | AS12 | *Serratia* sp. AS12 | COMPLETE | 1 | 5443009 | 55.96144 | 4974 | 22 | 87 |
| 3 | GCF_000214235.1 | AS9 | *Serratia* plymuthica AS9 | COMPLETE | 1 | 5442880 | 55.96115 | 4975 | 22 | 87 |
| 4 | GCF_000214805.1 | AS13 | *Serratia* sp. AS13 | COMPLETE | 1 | 5442549 | 55.96058 | 4975 | 22 | 87 |
| 5 | GCF_000330865.1 | FGI94 | *Serratia marcescens* FGI94 | COMPLETE | 1 | 4858216 | 58.86171 | 4363 | 22 | 83 |
| 6 | GCF_000336425.1 | WW4 | *Serratia marcescens* WW4 | COMPLETE | 2 | 5244703 | 59.54707 | 4831 | 22 | 83 |
| 7 | GCF_000478545.1 | RVH1 | *Serratia plymuthica* RVH1 | COMPLETE | 1 | 5514320 | 56.19833 | 5029 | 21 | 84 |
| 8 | GCF_000176835.2 | 4Rx13 | *Serratia plymuthica* 4Rx13 | COMPLETE | 2 | 5403731 | 56.1339 | 4913 | 22 | 82 |
| 9 | GCF_000438825.1 | S13 | *Serratia plymuthica* S13 | COMPLETE | 1 | 5467306 | 56.19532 | 4963 | 22 | 86 |
| 10 | GCF_000422085.1 | ATCC 27592(Type) | *Serratia liquefaciens* ATCC 27592 | COMPLETE | 2 | 5282719 | 55.3295 | 4898 | 24 | 86 |
| 11 | GCF_000513215.1 | Db11 | *Serratia marcescens* subsp. *marcescens* Db11 | COMPLETE | 1 | 5113802 | 59.51163 | 4676 | 22 | 88 |
| 12 | GCF_000828775.1 | SM39 | *Serratia marcescens* SM39 | COMPLETE | 3 | 5326023 | 59.74242 | 4892 | 22 | 88 |
| 13 | GCF_000695995.1 | FS14 | *Serratia* sp. FS14 | COMPLETE | 1 | 5249875 | 59.46231 | 4779 | 21 | 92 |
| 14 | GCF_000747565.1 | SCBI | *Serratia* sp. SCBI | COMPLETE | 2 | 5101896 | 59.5932 | 4714 | 21 | 82 |
| 15 | GCF_000975245.1 | HUMV-21 | *Serratia liquefaciens* | COMPLETE | 1 | 5326657 | 55.19101 | 4911 | 22 | 86 |
| 16 | GCF_001006005.1 | DSM 4576(Type) | *Serratia fonticola* | COMPLETE | 1 | 6000511 | 53.6128 | 5354 | 22 | 84 |
| 17 | GCF_001022215.1 | CAV1492 | *Serratia marcescens* | COMPLETE | 6 | 5828402 | 58.61262 | 5397 | 22 | 88 |
| 18 | GCA_000648575.1 | V4 | *Serratia plymuthica* | CHROMOSOME | 1 | 5513353 | 56.19424 | 5073 | 3 | 65 |
| 19 | GCF_000988045.1 | Lr5/4 | *Serratia ureilytica* | CONTIG | 1 | 5391360 | 59.18134 | 4984 | 22 | 90 |
| 20 | GCF_000521925.1 | BIDMC 44 | *Serratia marcescens* BIDMC 44 | SCAFFOLD | 2 | 5341662 | 59.43478 | 4829 | 22 | 86 |
| 21 | GCF_000347995.1 | S4 | *Serratia* sp. S4 | SCAFFOLD | 2 | 5454741 | 54.9871 | 4950 | 23 | 83 |
| 22 | GCF_000783915.1 | FDA_MicroDB_65 | *Serratia marcescens* | CONTIG | 1 | 5248255 | 59.87712 | 4780 | 22 | 88 |
| 23 | GCA_001034405.1 | UCI88 | *Serratia marcescens* | SCAFFOLD | 1 | 5187497 | 59.68626 | 4712 | 22 | 83 |
| 24 | GCF_000743395.1 | CDC_813-60(Type) | *Serratia marcescens* | SCAFFOLD | 2 | 5131648 | 59.78364 | 4724 | 22 | 99 |
| 25 | GCF_000163595.1 | DSM 4582(Type) | *Serratia odorifera* DSM 4582 | SCAFFOLD | 33 | 5267389 | 54.57442 | 4901 | 3 | 70 |
| 26 | GCF_000783975.1 | FDAARGOS_62 | *Serratia marcescens* | CONTIG | 4 | 5136411 | 59.65679 | 4777 | 22 | 92 |
| 27 | GCA_001034375.1 | BWH56 | *Serratia marcescens* | SCAFFOLD | 10 | 5620047 | 58.35971 | 5174 | 22 | 84 |
| 28 | GCF_000738675.1 | DSM 21420(Type) | *Serratia nematodiphila* | CONTIG | 2 | 5224920 | 59.53609 | 4807 | 24 | 91 |
| 29 | GCF_000264275.1 | LCT-SM213 | *Serratia marcescens* LCT-SM213 | SCAFFOLD | 11 | 5067328 | 59.15194 | 4633 | 5 | 73 |
| 30 | GCF_000521905.1 | BIDMC 50 | *Serratia marcescens* BIDMC 50 | SCAFFOLD | 4 | 5120658 | 59.56225 | 4697 | 24 | 85 |
| 31 | GCF_000739215.1 | NGS-ED-1015 | *Serratia marcescens* | CONTIG | 7 | 5122566 | 59.91275 | 4684 | 8 | 78 |
| 32 | GCF_000699165.1 | FK01 | *Serratia liquefaciens* FK01 | SCAFFOLD | 28 | 5280113 | 55.74578 | 4851 | 9 | 80 |
| 33 | GCA_001030265.1 | BWH57 | *Serratia marcescens* | SCAFFOLD | 13 | 5659252 | 58.27719 | 5217 | 26 | 84 |
| 34 | GCA_001034395.1 | UCI87 | *Serratia marcescens* | SCAFFOLD | 12 | 5133128 | 59.51811 | 4729 | 15 | 85 |
| 35 | GCF_000805875.1 | RM66262 | *Serratia marcescens* | CONTIG | 19 | 4882257 | 60.05288 | 4443 | 9 | 79 |
| 36 | GCF_000442455.1 | LCT-SM166 | *Serratia marcescens* LCT-SM166 | SCAFFOLD | 13 | 5070021 | 59.12378 | 4637 | 9 | 71 |
| 37 | GCA_001051865.1 | AH0650_Sm1 | *Serratia marcescens* subsp. *marcescens* | CONTIG | 20 | 5201657 | 59.52695 | 4751 | 24 | 92 |
| 38 | GCA_000783615.1 | FDAARGOS_79 | *Serratia marcescens* | CONTIG | 14 | 5381785 | 59.70844 | 5543 | 31 | 88 |
| 39 | GCF_001007555.1 | 90-166 | *Serratia marcescens* | CONTIG | 63 | 5484396 | 59.11298 | 5087 | 10 | 85 |
| 40 | GCF_000633715.1 | BIDMC 80 | *Serratia marcescens* BIDMC 80 | SCAFFOLD | 37 | 5274231 | 59.69994 | 4861 | 6 | 79 |
| 41 | GCF_000633555.1 | PH1a | *Serratia marcescens* PH1a | CONTIG | 56 | 5243840 | 59.2326 | 4894 | 6 | 81 |
| 42 | GCF_000633335.1 | H1q | *Serratia marcescens* H1q | CONTIG | 60 | 5243094 | 59.23398 | 4895 | 5 | 81 |
| 43 | GCF_000477615.1 | AU-AP2C | *Serratia fonticola* AU-AP2C | CONTIG | 34 | 4999819 | 53.75147 | 4606 | 4 | 68 |
| 44 | GCA_001063375.1 | 420_SMAR | *Serratia marcescens* | CONTIG | 37 | 5206386 | 59.43966 | 4789 | 9 | 74 |
| 45 | GCA_001064975.1 | 454_SMAR | *Serratia marcescens* | CONTIG | 33 | 5209606 | 59.44645 | 4797 | 13 | 77 |
| 46 | GCF_000468075.1 | AU-P3(3) | *Serratia fonticola* AU-P3(3) | CONTIG | 44 | 5023127 | 53.75381 | 4704 | 5 | 65 |
| 47 | GCF_000821185.1 | SAf | *Serratia symbiotica* | CONTIG | 32 | 3581747 | 52.07998 | 3597 | 35 | 74 |
| 48 | GCF_000442375.1 | LCT-SM262 | *Serratia marcescens* LCT-SM262 | SCAFFOLD | 55 | 5160155 | 55.99328 | 4740 | 5 | 80 |
| 49 | GCF_000633695.1 | BIDMC 81 | *Serratia marcescens* BIDMC 81 | SCAFFOLD | 33 | 5046223 | 59.66978 | 4655 | 11 | 65 |
| 50 | GCA_001062925.1 | 20_SPLY | *Serratia liquefaciens* | CONTIG | 40 | 5164504 | 55.4322 | 4746 | 8 | 68 |
| 51 | GCF_000734475.1 | YDC563 | *Serratia marcescens* | CONTIG | 54 | 5073647 | 59.97004 | 4629 | 5 | 74 |
| 52 | GCA_001064715.1 | 370_SMAR | *Serratia marcescens* | CONTIG | 52 | 5196323 | 59.46287 | 4783 | 5 | 44 |
| 53 | GCF_000342205.1 | VGH107 | *Serratia marcescens* VGH107 | CONTIG | 83 | 5519124 | 59.46411 | 5157 | 6 | 85 |
| 54 | GCF_000292365.1 | W2.3 | *Serratia marcescens* W2.3 | CONTIG | 72 | 5231794 | 59.23888 | 4891 | 5 | 74 |
| 55 | GCA_001063175.1 | 286_SMAR | *Serratia marcescens* | CONTIG | 49 | 5120464 | 59.7706 | 4754 | 4 | 66 |
| 56 | GCA_001064335.1 | 287_SMAR | *Serratia marcescens* | CONTIG | 54 | 5122819 | 59.76729 | 4755 | 5 | 68 |
| 57 | GCA_001063145.1 | 280_SMAR | *Serratia marcescens* | CONTIG | 55 | 5113427 | 59.76651 | 4748 | 3 | 66 |
| 58 | GCF_000743355.1 | Ag1 | *Serratia* sp. Ag1 | CONTIG | 110 | 5347575 | 52.2653 | 4965 | 13 | 72 |
| 59 | GCF_000186485.1 | Tucson | *Serratia symbiotica* str. Tucson | SCAFFOLD | 388 | 2789218 | 47.94917 | 3334 | 22 | 51 |
| 60 | GCF_000465615.1 | EGD-HP20 | *Serratia marcescens* EGD-HP20 | CONTIG | 62 | 5083006 | 59.76031 | 4705 | 5 | 81 |
| 61 | GCA_001063125.1 | 276_SMAR | *Serratia marcescens* | CONTIG | 108 | 5163596 | 59.56804 | 4792 | 9 | 64 |
| 62 | GCA_001076625.1 | 294_SMAR | *Serratia marcescens* | CONTIG | 75 | 5113656 | 59.76996 | 4751 | 5 | 46 |
| 63 | GCA_001064855.1 | 398_SMAR | *Serratia marcescens* | CONTIG | 64 | 5198466 | 59.45135 | 4794 | 5 | 50 |
| 64 | GCA_001065325.1 | 532_SMAR | *Serratia marcescens* | CONTIG | 89 | 5099230 | 59.71996 | 4793 | 3 | 35 |
| 65 | GCA_001064345.1 | 290_SMAR | *Serratia marcescens* | CONTIG | 61 | 5115483 | 59.77164 | 4746 | 5 | 54 |
| 66 | GCA_001062285.1 | 1198.rep1_SMAR | *Serratia marcescens* | CONTIG | 77 | 5159990 | 59.93099 | 4707 | 2 | 43 |
| 67 | GCA_001063325.1 | 410_SMAR | *Serratia marcescens* | CONTIG | 74 | 5202811 | 59.44312 | 4810 | 11 | 52 |
| 68 | GCF_000738535.1 | MCB | *Serratia marcescens* | CONTIG | 103 | 5298782 | 59.09839 | 4912 | 14 | 77 |
| 69 | GCF_000300895.1 | A30 | *Serratia plymuthica* A30 | CONTIG | 78 | 5554023 | 56.10276 | 5179 | 4 | 65 |
| 70 | GCF_000751195.1 | | *Serratia marcescens* | CONTIG | 66 | 5100993 | 59.81294 | 4746 | 15 | 84 |
| 71 | GCF_000743365.1 | Ag2 | *Serratia* sp. Ag2 | CONTIG | 109 | 5324817 | 52.27147 | 4928 | 16 | 69 |
| 72 | GCF_000418935.1 | AB42556419-isolate1 | *Serratia marcescens* AB42556419-isolate1 | SCAFFOLD | 117 | 5334871 | 59.19279 | 5008 | 4 | 64 |
| 73 | GCF_000418835.1 | MC6001 | *Serratia marcescens* MC6001 | SCAFFOLD | 131 | 5367622 | 59.08902 | 5046 | 7 | 61 |
| 74 | GCF_000418895.1 | MC459 | *Serratia marcescens* MC459 | SCAFFOLD | 129 | 5373060 | 59.08992 | 5058 | 5 | 66 |
| 75 | GCA_001064835.1 | 395_SMAR | *Serratia marcescens* | CONTIG | 79 | 5199539 | 59.45808 | 4799 | 6 | 37 |
| 76 | GCA_001060625.1 | 1241_SMAR | *Serratia marcescens* | CONTIG | 117 | 5117553 | 59.3411 | 4732 | 6 | 29 |
| 77 | GCF_000418875.1 | MC460 | *Serratia marcescens* MC460 | SCAFFOLD | 139 | 5371197 | 59.0638 | 5053 | 8 | 62 |
| 78 | GCF_000418855.1 | MC6000 | *Serratia marcescens* MC6000 | SCAFFOLD | 134 | 5333162 | 59.19666 | 5015 | 7 | 61 |
| 79 | GCF_000418815.1 | MC620 | *Serratia marcescens* MC620 | SCAFFOLD | 138 | 5259807 | 59.18253 | 4923 | 6 | 68 |
| 80 | GCA_00106737 | 790_SMAR | *Serratia* | CONTIG | 159 | 568713 | 59.3041 | 5342 | 7 | 50 |
| | 5.1 | | *marcescens* | | | 3 | 5 | | | |
| 81 | GCA_001062235.1 | 1185_SMAR | *Serratia marcescens* | CONTIG | 139 | 5163463 | 59.57099 | 4837 | 7 | 35 |
| 82 | GCF_000734885.1 | A2 | *Serratia grimesii* | CONTIG | 120 | 5133068 | 52.842 | 5013 | 14 | 69 |
| 83 | GCF_000496755.2 | DD3 | *Serratia* sp. DD3 | CONTIG | 123 | 5273532 | 49.16261 | 4659 | 23 | 73 |
| 84 | GCA_001060655.1 | 1242_SMAR | *Serratia marcescens* | CONTIG | 144 | 5113929 | 59.34494 | 4748 | 5 | 29 |
| 85 | GCA_001067015.1 | 706_SMAR | Serratia marcescens | CONTIG | 125 | 5106204 | 60.16313 | 4669 | 2 | 32 |
| 86 | GCA_001061195.1 | 1198.rep2_SMAR | *Serratia marcescens* | CONTIG | 135 | 5126237 | 59.92799 | 4727 | 5 | 30 |
| 87 | GCF_000476315.1 | UTAD54 | *Serratia fonticola* UTAD54 | CONTIG | 133 | 5953423 | 53.53411 | 5552 | 4 | 68 |
| 88 | GCA_000336365.2 | C-1 | *Serratia* sp. C-1 | SCAFFOLD | 133 | 5651614 | 56.16006 | 5221 | 3 | 63 |
| 89 | GCF_000418915.1 | MC458 | *Serratia marcescens* MC458 | SCAFFOLD | 156 | 5402091 | 59.14523 | 5080 | 6 | 67 |
| 90 | GCF_000261045.1 | PRI-2C | *Serratia plymuthica* PRI-2C | CONTIG | 104 | 3919163 | 55.785 | 3668 | 3 | 30 |
| 91 | GCA_001064455.1 | 311_SMAR | *Serratia marcescens* | CONTIG | 164 | 5095396 | 59.80395 | 4757 | 7 | 32 |
| 92 | GCA_001060585.1 | 1218_SMAR | *Serratia marcescens* | CONTIG | 154 | 5126894 | 59.92966 | 4746 | 3 | 31 |
| 93 | GCA_001061145.1 | 1186_SMAR | *Serratia marcescens* | CONTIG | 204 | 5152789 | 59.57684 | 4870 | 7 | 31 |
| 94 | GCA_001068085.1 | 907_SMAR | *Serratia marcescens* | CONTIG | 185 | 4920623 | 59.62997 | 4630 | 3 | 28 |
| 95 | GCA_001061225.1 | 1219_SMAR | *Serratia marcescens* | CONTIG | 156 | 5137374 | 59.91076 | 4768 | 5 | 33 |
| 96 | GCA_001065935.1 | 684_SMAR | *Serratia marcescens* | CONTIG | 239 | 5472637 | 59.00724 | 5185 | 2 | 32 |
| 97 | GCF_000633315.1 | H1n | *Serratia* sp. H1n | CONTIG | 247 | 6123355 | 53.83338 | 5773 | 9 | 74 |
| 98 | GCF_000633355.1 | H1w | *Serratia* sp. H1w | CONTIG | 240 | 6126027 | 53.83577 | 5770 | 5 | 74 |
| 99 | GCA_001066945.1 | 698_SMAR | *Serratia marcescens* | CONTIG | 193 | 5213052 | 59.99349 | 4824 | 2 | 29 |
| 100 | GCA_001064725.1 | 374_SMAR | *Serratia marcescens* | CONTIG | 197 | 5185093 | 59.47089 | 4872 | 4 | 32 |

Table 2. General statistical values (mean, median, standard deviation) for genome size and other properties for the studied genus. Intergenic regions are spacers or regions between genes, shorter intergenic regions mean more compact arrangement of genes and hence, genome.

| Property | Mean | Median | SD |
|---|---|---|---|
| Genome size (Mbp) | 5.23 | 5.21 | 0.4 |
| GC content (%) | 57.95 | 59.27 | 2.62 |
| No. of CDS | 4852.01 | 4817 | 335.95 |
| Mean length of CDS | 940.4 | 949.08 | 38.08 |
| Mean length of intergenic region | 149.25 | 146.11 | 13.5 |

Table 3. Regression analysis of genome properties. For regression analysis of genome size to the number of CDSs among 100 *Serratia* strains, number of CDSs within this species were positively correlated to genome size ($r^2$=0.95) and the coefficient of determination ($R^2$=0.9) meant that 90% of data points supported the predicted regression line (*y* = 719 + 790x).

| Property | Genome size (Mbp) | GC content (%) | No. of CDS | Mean length of CDS | Mean length of intergenic region |
|---|---|---|---|---|---|
| Genome size (Mbp) | | Y=4+0.021x<br>$R^2$=0.02<br>$r^2$=0.13 | Y=-0.31+0.0011x<br>$R^2$=0.9<br>$r^2$=0.95 | Y=-0.79+0.0064<br>$R^2$=0.36<br>$r^2$=0.6 | Y=6.7-0.01x<br>$R^2$=0.11<br>$r^2$=-0.33 |
| GC content (%) | Y= 53+0.87x<br>$R^2$=0.02<br>$r^2$=0.13 | | Y=55+0.00054x<br>$R^2$=0<br>$r^2$=0.07 | Y=32+0.026<br>$R^2$=0.16<br>$r^2$=0.4 | Y=7+0.09x<br>$R^2$=0.21<br>$r^2$=-0.46 |
| No. of CDS | Y=719+790x<br>$R^2$=0.9<br>$r^2$=0.95 | Y=55+0.00054x<br>$R^2$=0<br>$r^2$=0.07 | | Y=1880+3.2x<br>$R^2$=0.13<br>$r^2$=0.36 | Y=5936-7.3x<br>$R^2$=0.09<br>$r^2$=-0.2 |
| Mean length of CDS | Y=643+57x<br>$R^2$=0.36<br>$r^2$=0.6 | Y=32+0.028x<br>$R^2$=0.16<br>$r^2$=0.4 | Y=743+0.041x<br>$R^2$=0.13<br>$r^2$=0.36 | | Y=1192-1.7x<br>$R^2$=0.26<br>$r^2$=-0.6 |
| Mean length of intergenic region | Y=207-11x<br>$R^2$=0.11<br>$r^2$=-0.33 | Y=74+0.09x<br>$R^2$=0.21<br>$r^2$=0.46 | Y=206-0.012x<br>$R^2$=0.09<br>$r^2$=-0.29 | Y=349-0.21x<br>$R^2$=0.36<br>$r^2$=-9.6 | |

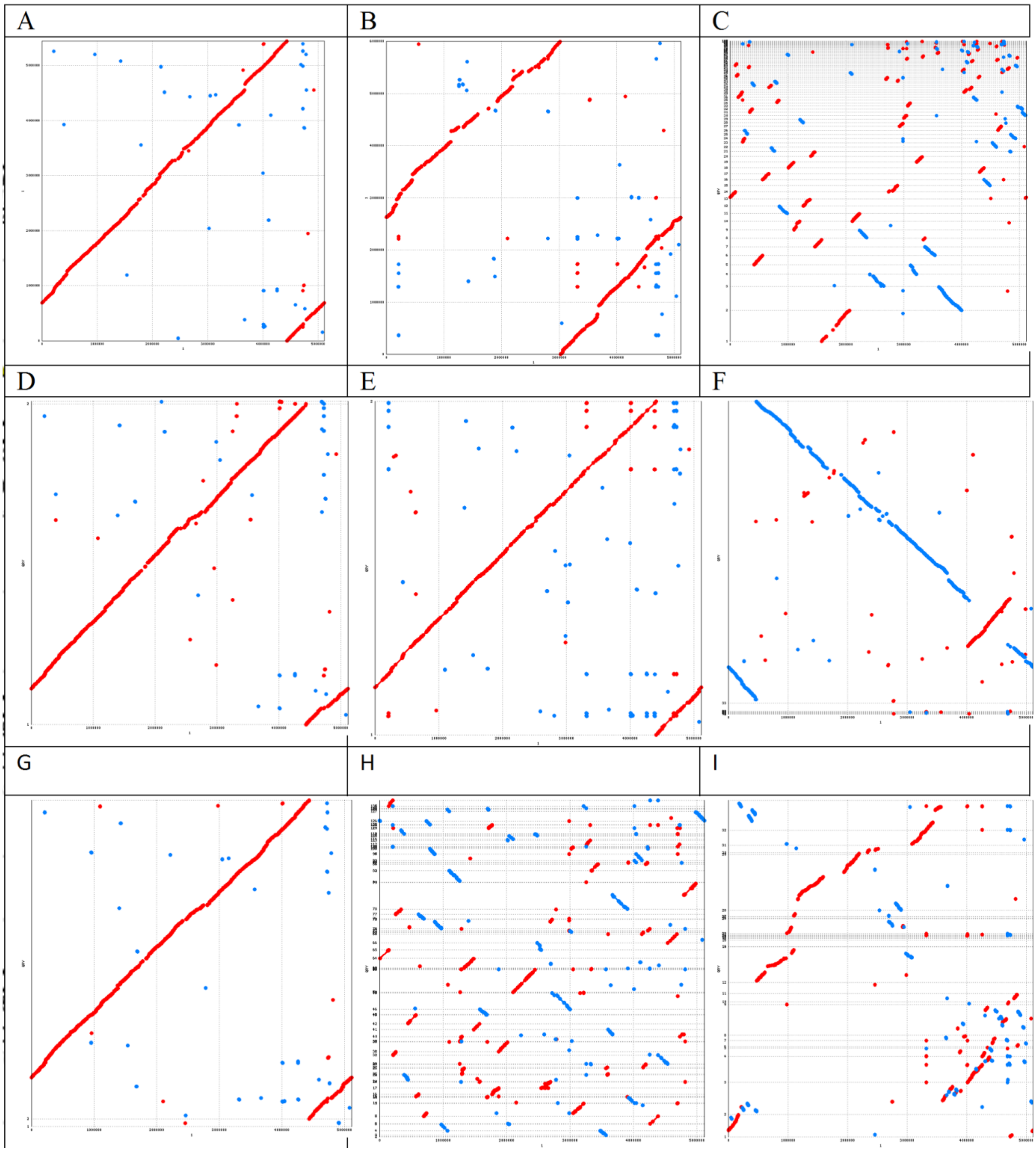

Fig. 1.NUCMER alignment plot of studied 10 species, classified according to ANI scheme and including (A) *Serratia plymuthica* AS12 , (B) *Serratia fonticola* DSM 4576, (C) *Serratia grinesii* A2, (D) *Serratia liquefaciens* ATCC 27592, (E) *Serratia nematodiphila* WW4, (F) *Serratia odorifera* DSM 4582, (G) *Serratia proteamaculans* S4, (H) *Serratia* subsp sakuensis and (I) *Serratia symbiotica* SAf with respect to the reference genome specie *Serratia marcescens* subspecie *marcescens* Db11. Rearrangement of genomic regions was observed in the genomes, in term of collinearity and their localization on the negative or positive strand.

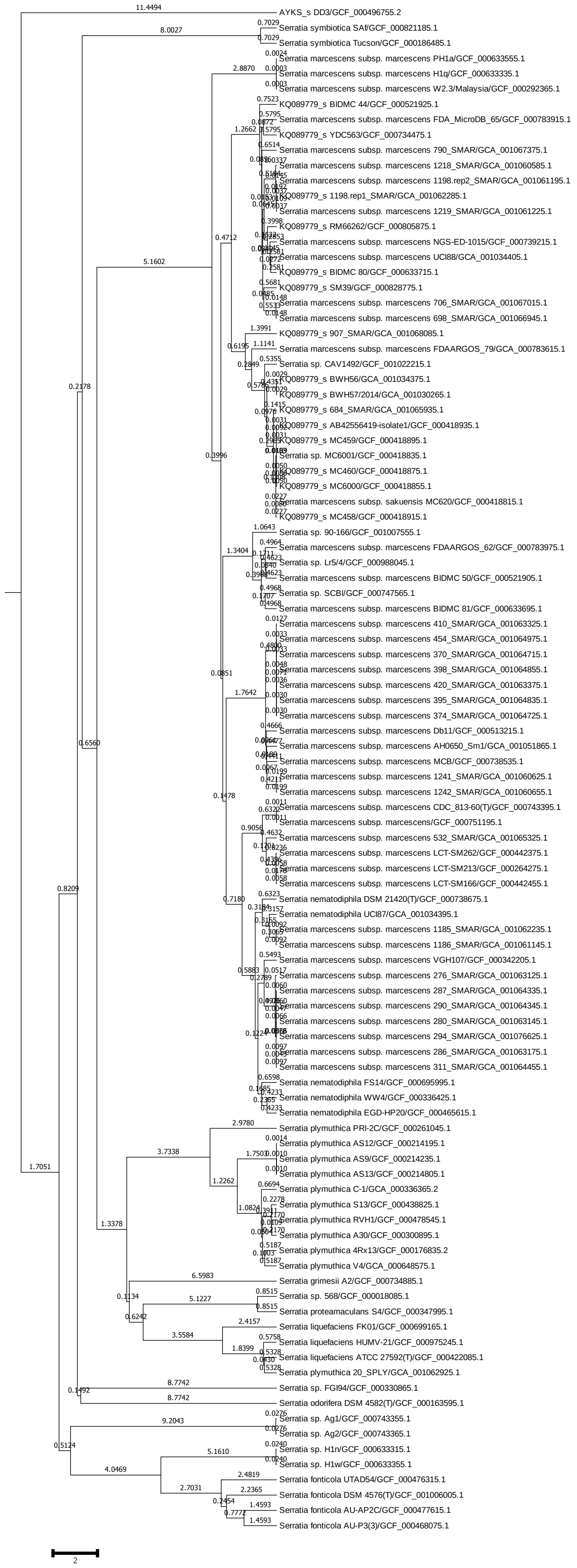

Fig. 2. Phylogeny of studied genomes based on ANI.

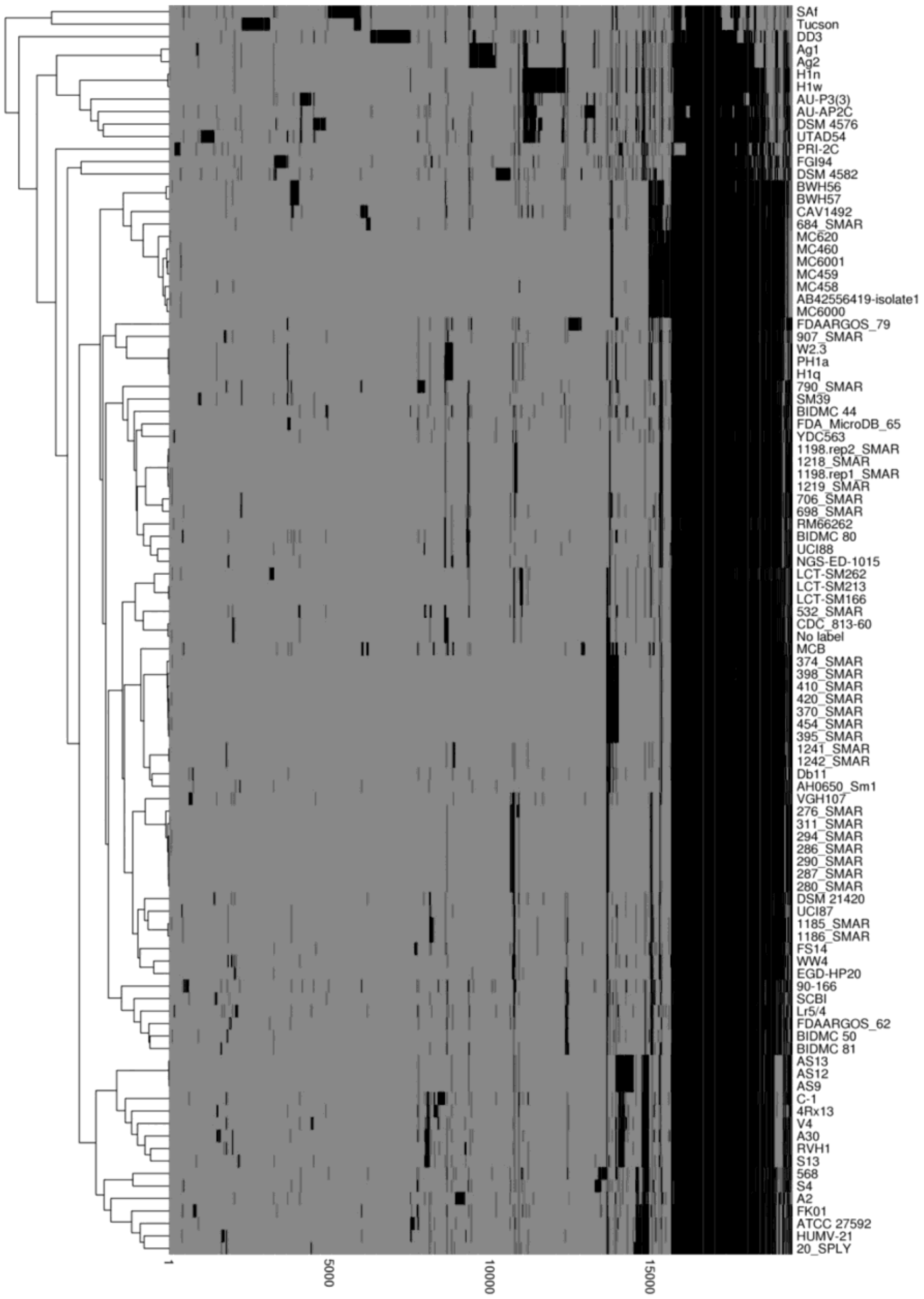

Fig. 3. Gene presence/absence matrix of all CDSs of 100 Serratia genomes. Black depicts presence while gray shows absence of a gene. Specie names are shown on the right side. The strains clustered on the on top i.e. SAf, Tucson, DD3, Ag1, Ag2, H1n and h1w show presence of significant number of genes that are absent in the other strains.

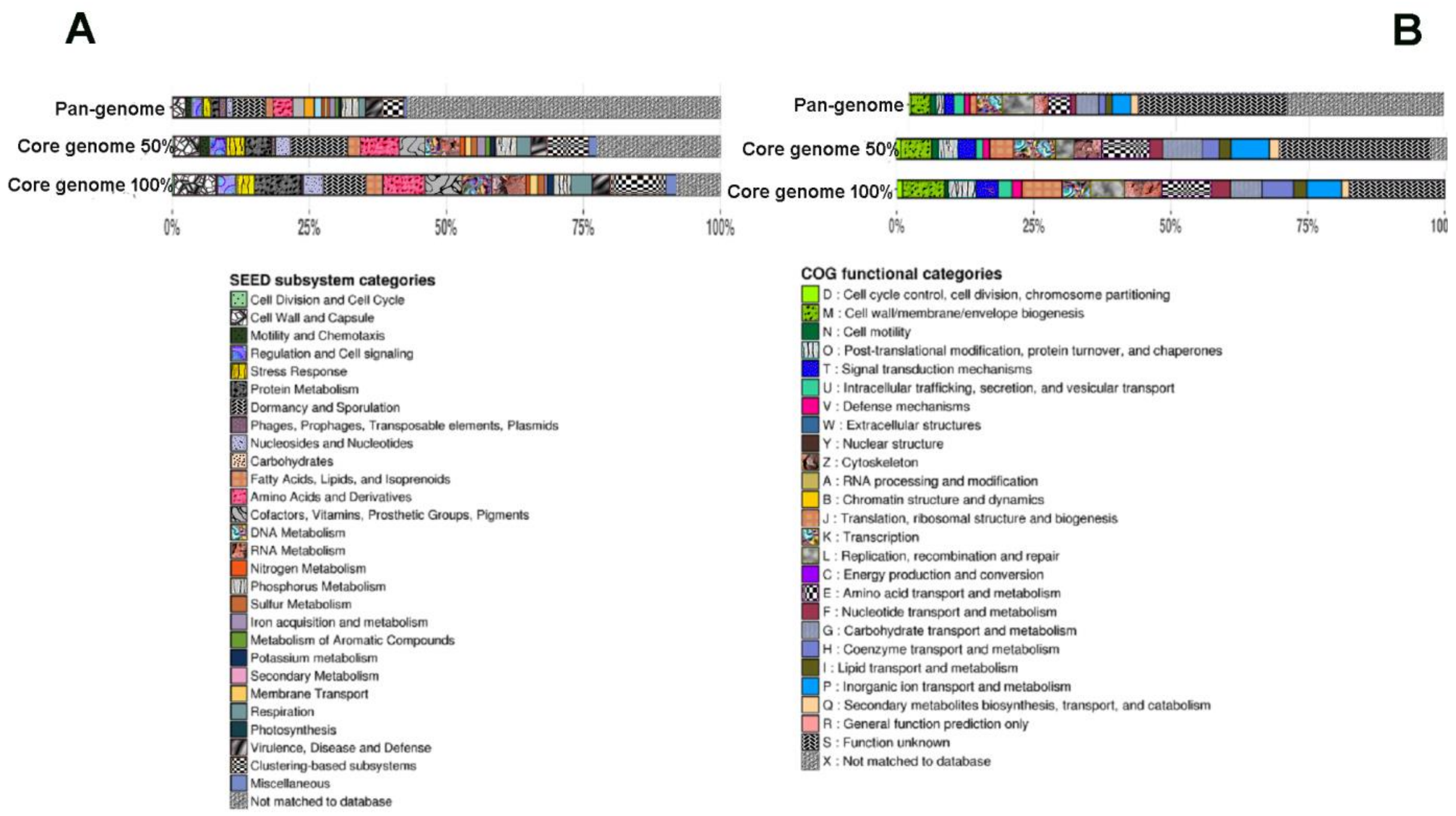

Fig. 4. Functional categories according to seed scheme. (B) Functional categories by COG scheme. Cut-off of 50% and 100% for core genome is shown.

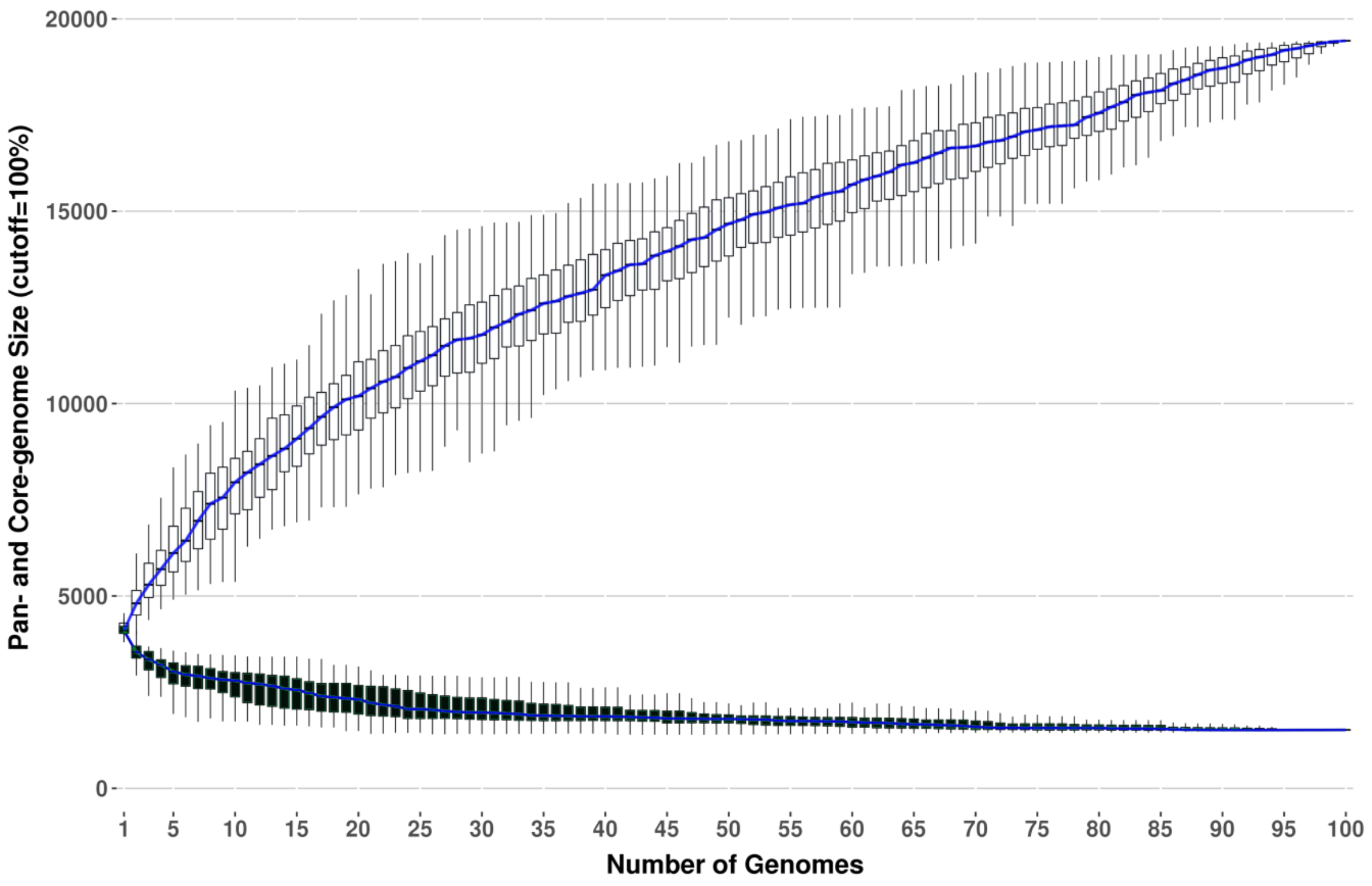

Fig. 5. Pan (top) and core-genome (bottom) curves, calculated from POGs of 100 *Serratia* genomes. Core genome cut-off is 100% in this figure.